\documentclass[runningheads]{svmult}

\usepackage{makeidx}   
\usepackage{graphicx}  
\usepackage{subeqnar}  
\usepackage{multicol}  
\usepackage{physprbb}  
\makeindex             

\usepackage{amsmath,amssymb} 

\def\mh{M_\bullet}
\def\msun{M_\odot}

\begin{document}
\title*{Interaction of Supermassive Black Holes with their Stellar and Dark Matter Environments}
\toctitle{Interaction of Supermassive Black Holes with their Stellar and Dark Matter Environments}
%
%
\titlerunning{Interaction of Black Holes with their Environment}
%
\author{David Merritt}
\authorrunning{David Merritt}
%
%
\institute{Rochester Institute of Technology, Rochester, NY, USA}

\maketitle              

\section{Introduction}

While supermassive black holes probably gained most of their
mass via accretion of gas,
the galactic nuclei in which they are currently situated
are dominated by stars.
This article reviews recent theoretical work on the
interaction between black holes and their stellar environment,
and highlights ways in which the observed structure
of galactic nuclei can be used to constrain the formation
history of black holes.
Nuclei may also contain dark matter, and 
the possibility of detecting supersymmetric particles via
annihilation radiation from the Galactic center
has generated some interest\cite{hooper-04}.
The evolution of the dark matter distribution in the 
presence of a black hole in a stellar nucleus is also 
discussed.

\section{Matter Distribution around Black Holes}

\begin{figure}[t]
\begin{center}
\includegraphics[width=.90\textwidth]{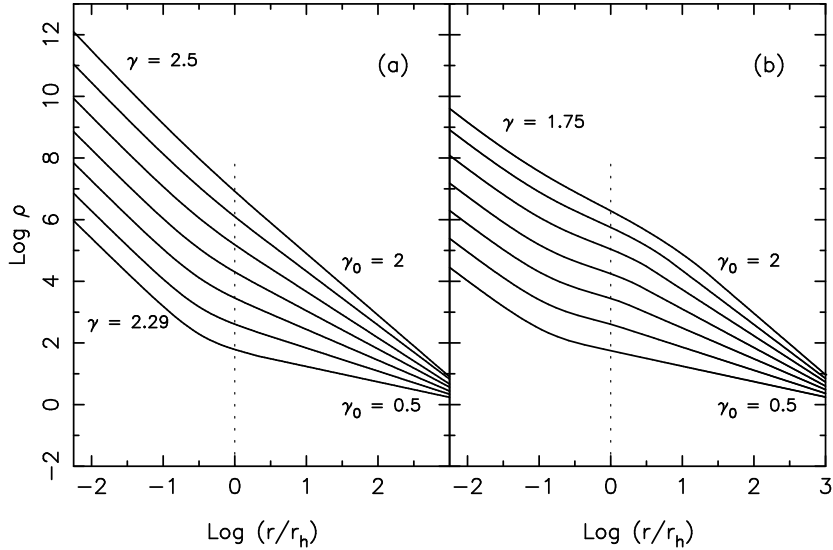}
\end{center}
\caption[]{(a) ``Collisionless'' density profiles,
resulting from growth of a black hole at a fixed location.
Initial profiles were power laws, $\rho\propto r^{-\gamma_0}$,
with $\gamma_0$ increasing upwards in steps of $0.25$.
The radial scale is normalized to $r_h$ as defined in the 
initial (pre-black-hole) galaxy.
(b) ``Collisional'' density profiles,
computed as steady-state solutions to the Fokker-Planck
equation.
The density inside $\sim 0.1 r_h$ satisfies $\rho\propto r^{-7/4}$.}
\label{fig:grow}
\end{figure}

Figure~\ref{fig:grow} shows two natural models for the
evolution of the stellar density around a black hole.
If the star-star relaxation time is long,
the form of $\rho(r)$ will still reflect the way the nucleus
formed.
The simplest possible picture is growth of a black
hole at a fixed location in space due to some spherically-symmetric
accretion process.
As the black hole grows it pulls in stars.
If the growth time is long (``adiabatic'') compared 
with orbital periods, the change in $\rho(r)$ is unique and 
straightforward to compute \cite{peebles-72,young-80}.
Figure~\ref{fig:grow}a shows the result if the initial
density is a power law, $\rho\propto r^{-\gamma_0}$.
The final profile is well described as two power laws
joined at a radius $r_{\rm cusp}$ where
\begin{equation}
r_{\rm cusp}=\alpha r_h,\ \ \ \ 0.19\lesssim\alpha\lesssim 0.22, \ \ \ \ 
0.5\le\gamma_0\le 1.5
\end{equation}
and $r_h$ is the radius containing a mass 
in stars equal to twice the final black hole mass.
The final density within $r_{\rm cusp}$ is
\begin{equation}
\rho\propto r^{-\gamma},\ \ \ \  2.29\le\gamma\le 2.5.
\end{equation}
This result is insensitive to the geometry
(flattened, triaxial, etc.) and degree of rotation of the nucleus.

If the nucleus is older than roughly one relaxation time,
a different sort of mass distribution is set up around the
black hole, corresponding to a steady-state solution of
the Fokker-Planck equation.
This steady state has a nearly zero net 
flux of stars with respect to energy and the density profile
is given uniquely as $\rho\propto r^{-7/4}$ near the black 
hole~\cite{sl-76,bw-76}.
Figure~\ref{fig:grow}b shows the result, again
assuming various power-law slopes for the initial $\rho(r)$.
The $r^{-7/4}$ dependence is only reached at
$r\lesssim r_{\rm cusp}\approx 0.1 r_h$.

The condition for the appearance of a collisional, $\gamma=7/4$
cusp is that the relaxation time,
\begin{equation}
T_r(r) = {\sqrt{2}\sigma(r)^3\over\pi G^2m\rho(r)\log\Lambda},
\label{eq:tr}
\end{equation}
measured at $\sim r_h$ 
be less than the age of the nucleus \cite{bw-76,preto-04}.
Figure~\ref{fig:tscales} shows $T_r(r_h)$ for a sample
of elliptical galaxies drawn from Faber et al. \cite{faber-97}.
Only a few galaxies, all low luminosity ellipticals,
have $T_r<10^{10}$ yr.
Two examples are well known: the Milky Way bulge,
which has $T_r\approx 2\times 10^9$ yr (assuming $m=M_\odot$),
and M32, for which $T_r\approx 3\times 10^9$ yr.
Both galaxies should have collisionally relaxed nuclei.
In the Milky Way, $r_h\approx 1.7$ pc $\approx 40''$ \cite{genzel-03}
and $r_{\rm cusp}\approx 4''$.
The stellar density profile has been determined into $\sim 0.1''$;
it is a broken power law with $\gamma\approx 1.4$ inside
of $\sim 10''$ and $\gamma\approx 2.0$ outside \cite{genzel-03}.
The break radius is reasonably close to the predicted value of
$r_{\rm cusp}$
but the interior slope is significantly flatter than $7/4$.
In the case of M32, $r_h\approx 2.9$ pc $\approx 0.9''$
and $r_{\rm cusp}\approx 0.1''$ \cite{lauer-98}.
The stellar luminosity density has been determined into
only $\sim 0.15''$, not quite far enough to test for the
existence of a collisional cusp.
The situation is worse for more distant galaxies that might
harbor collisionally relaxed nuclei, making the Milky Way nucleus
the only system in which the theory can currently be tested.
The fact that $\gamma<7/4$ in the Milky Way nucleus 
may be due to mass segregation which produces a flatter density profile
for the less-massive stars \cite{bw-77,mcd-91}.

\begin{figure}[t]
\begin{center}
\includegraphics[width=.65\textwidth]{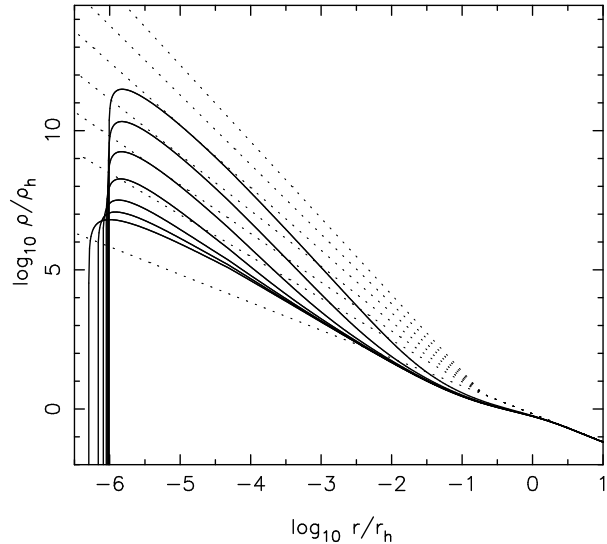}
\end{center}
\caption[]{Dark matter density at $t=0$
(dotted lines) and $10^{10}$ yr (solid lines)
after heating by stars in the Galactic bulge \cite{merritt-04}.}
\label{fig:dm}
\end{figure}

Particle dark matter, if present in galactic
nuclei, would respond to the growth of the black
hole according to the ``collisionless'' model.
Its density near the black hole could therefore
be very high, resulting in a substantial
rate of self-annihilations if the particles
are supersymmetric neutralinos \cite{gs-99}.
The dark matter distribution would also evolve due to scattering off
of stars, on a time scale close to $T_r$,
the star-star relaxation time \cite{ilyin-03,merritt-04,gp-04}.
In effect, the stars act as a heat source for the dark matter.
Figure~\ref{fig:dm} illustrates this via time-dependent solutions
to the Fokker-Planck equation.
Given the appropriate normalization with respect to
the dark matter density near the solar circle,
this nuclear profile implies a potentially observable flux
of annihilation products \cite{merritt-04}.

\section{Feeding Rates}

In a spherical galaxy, stars or dark matter particles with angular
momenta less than $J_{lc}\approx \sqrt{2G\mh r_t}$ will pass within
a distance $r_t$ of the center of the black hole.
The ``loss cone'' is that set of orbits satisfying $J\le J_{lc}(E)$.
If we imagine that the black hole is embedded in a steady-state
galaxy populated by stars with $J\le J_{lc}$, 
stars on loss cone orbits will be consumed in one orbital
period or less, giving a transitory feeding rate of 
$\sim 4\pi^2f(E,0)J_{lc}^2(E)$ stars per unit energy per unit time.
This is sometimes called the ``full loss cone'' consumption rate
and it is an effective upper limit to how quickly stars can
be supplied to the black hole.
For instance, in a $\rho\sim r^{-2}$ nucleus,
the full loss cone feeding rate due to all stars with $E>\Phi(r_h)$
is $\sim \sigma^5r_t/G^2\mh$ or $\sim \sigma^5/Gc^2$ when
$r_t\approx r_s\equiv 2GM_\bullet/\sigma^2$.
This is a high enough rate to grow black holes of the
correct size after $10^{10}$ yr, i.e. 
$\sim 10^{-2}\msun$ yr$^{-1}$ for $\sigma\approx 200$ km s$^{-1}$.

Full loss cones have been postulated from time to time,
perhaps first by Hills~\cite{hills-75}, later by P. Young
and co-authors (for the ``black-tide'' 
model of quasar fueling)~\cite{young-77,ysw-77}, 
and most recently by Zhao, Haehnelt \& Rees~\cite{zhao-02},
who noted that the $\sigma^5$ scaling is consistent with the
$M_\bullet-\sigma$ relation.
But while transient loss cone refilling during, say,
galaxy mergers is imaginable, it is hard to see how 
a galaxy could arrange for the $10^5$ or more refilling
events that would be needed to keep a loss cone
continuously filled.

Instead, the loss cone is probably usually empty at radii $r<r_{crit}$,
where $r_{crit}$ is the radius at which the scattering time
into the loss cone equals the orbital period.
Below $r_{crit}$, stars are lost the moment they enter the loss cone,
while above $r_{crit}$, stars may scatter in and out of the loss
cone in a single orbit.
For $\mh\approx 10^7\msun$,
$r_{crit}\approx r_h$,
and $r_{crit}<r_h$ at larger $\mh$~\cite{fr-76}.
The feeding rate due to stars inside $r_{crit}$
is $\sim M(r_{crit})/T_r(r_{crit})$
since in one relaxation time a star's angular momentum
can change by of order itself.
Outside $r_{crit}$, the loss cone is effectively
full but the feeding rate drops off rapidly with
radius since the capture sphere subtends such a small
angle (the ``pinhole'' regime).
One  finds in a careful calculation that the
flux into the black hole is strongly peaked at
energies near $\Phi(r_{crit})$~\cite{ls-77,ck-78}.

\begin{figure}[t]
\begin{center}
\includegraphics[width=.65\textwidth]{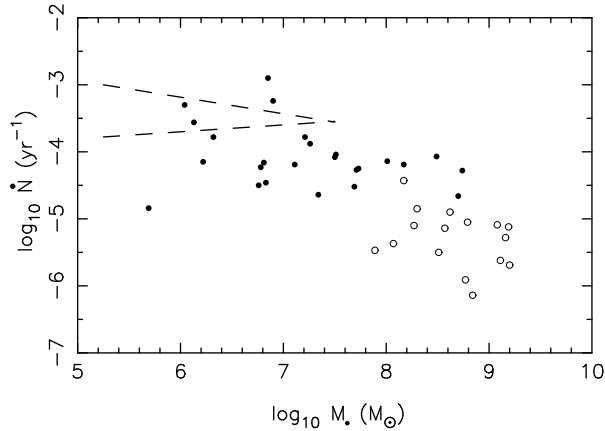}
\end{center}
\caption[]{Stellar tidal disruption rates in a sample of elliptical galaxies
computed using the Cohn-Kulsrud \cite{ck-78} loss cone boundary 
conditions \cite{wang-04}.
Open circles are ``core'' galaxies and filled circles are ``power-law''
galaxies.
Upper and lower dashed lines are disruption rates in power-law
nuclei with $\rho\propto r^{-2}$ and $r^{-1.5}$ respectively.}
\label{fig:ndot}
\end{figure}

Figure~\ref{fig:ndot} shows tidal disruption rates computed in a sample of
elliptical galaxies, with $r_t$ set to the tidal
 radius for solar-mass stars \cite{wang-04}.
(For $M_\bullet\gtrsim 10^8\msun$, $r_t<r_s$ and stars would be
swallowed whole.)
The numbers in this plot supersede the estimates
in~\cite{mt-99} which were based on the now-discredited
Magorrian et al.~\cite{magorrian-98} black hole masses.
Disruption rates reach $\sim 10^{-3}$ yr$^{-1}$ in the
faintest galaxies with verified black holes like M32.
The dashed lines in that figure are the predicted 
rates in a singular isothermal sphere nucleus:
\begin{equation}
\dot N \approx 2.5\times 10^{-3} {\rm yr}^{-1} 
\left({\sigma\over 100\ {\rm km\ s}^{-1}}\right)^{7/2}\left({\mh\over 10^6\msun}\right)^{-1} \propto M_\bullet^{-0.25}
\label{rate1}
\end{equation}
and in a nucleus with $\rho\propto r^{-3/2}$:
\begin{equation}
\dot N \approx 1.0\times 10^{-3} {\rm yr}^{-1} \left({\sigma\over 100\ {\rm km\ s}^{-1}}\right)^{21/5}\left({\mh\over 10^6\msun}\right)^{-4/5}
\propto M_\bullet^{0.10}
 \label{rate2}
\end{equation}
where the latter expressions in each case use the $\mh-\sigma$ 
relation \cite{mf-01}.
The tidal flaring rate is probably nearly independent
of $M_\bullet$ for $M_\bullet\lesssim 10^6M_\odot$,
and if dE galaxies contain black holes, they would
dominate the total flaring rate \cite{wang-04}.

A fraction $25\%-50\%$ of a tidally disrupted star is
expected to remain gravitationally bound to the black hole
\cite{ayal-00}.
If this is the case, and if disruption rates have remained nearly
constant over $10^{10}$ yr, 
Figure~\ref{fig:ndot} suggests that stellar
consumption could be a major contributor to the growth
of black holes with $M_\bullet\lesssim 10^7M_\odot$.

\begin{figure}[t]
\begin{center}
\includegraphics[width=.70\textwidth]{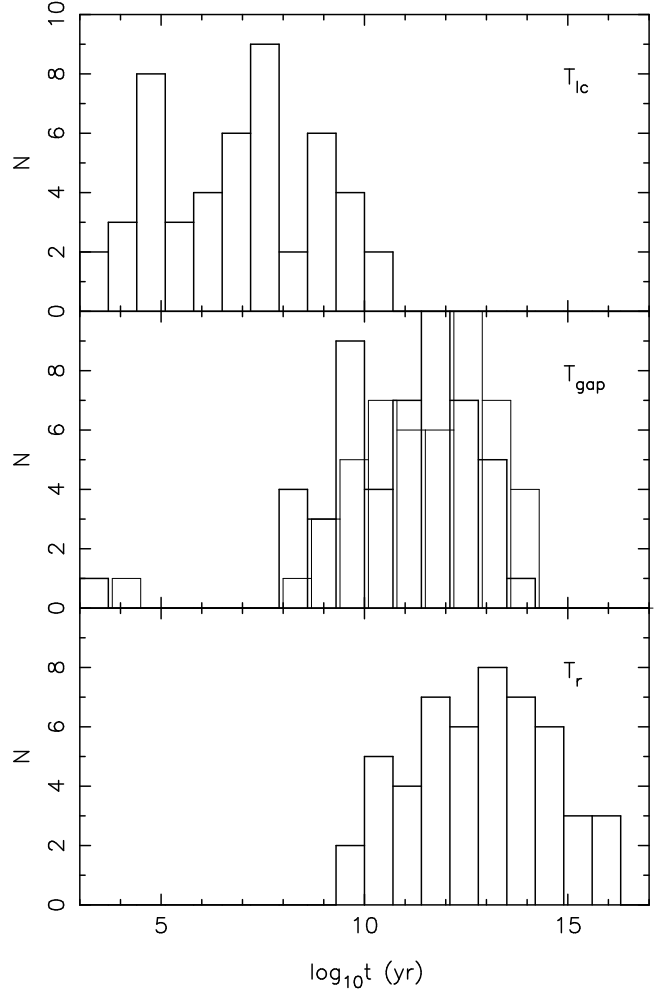}
\end{center}
\caption[]{Time scales $T_r$, $T_{lc}$ and $T_{gap}$ defined in the text,
for the sample of elliptical galaxies in Fig.~\ref{fig:ndot}.
Thick/thin histograms of $T_{gap}$ are for $m_2/m_1=(0.1,1)$,
where $m_2/m_1$ is the mass ratio of the binary black hole
that preceded the current, single black hole.}
\label{fig:tscales}
\end{figure}

The feeding rates estimated above were based on a theory
derived in the late 1970's to describe consumption rates of
stars by massive black holes at the centers of globular clusters
\cite{bw-76,ls-77,ck-78}.
Globular clusters are many relaxation times old,
and the assumption of a collisionally-relaxed steady state 
is built into the theory.
For instance, the Cohn-Kulsrud \cite{ck-78} loss cone
boundary conditions, used in Figure~\ref{fig:ndot},
are only valid in a collisionally relaxed system.
But relaxation times in galactic nuclei can be very long
and this fact requires a re-thinking of loss cone theory.
Figure \ref{fig:tscales} shows histograms of three time scales relevant
to feeding rates.
The bottom panel is the relaxation time $T_r$ (eq. \ref{eq:tr})
evaluated at the black hole's influence radius $r_h$.
$T_r$ exceeds $10^{10}$ yr in most galaxies.
The middle and upper panels show two time scales
more directly related to the establishment of a steady-state loss cone. 
The time scale for scattering of low angular momentum stars
into the loss cone is
\begin{eqnarray}
T_\theta(r) \approx \theta(r)^2T_r(r)
\end{eqnarray}
where $\theta(r)$ is the scattering angle as seen from 
a star at radius $r$;
the square-root dependence of $\theta$ on $T_\theta$
reflects the fact that entry into the loss cone is a diffusive process.
If the stellar phase space is initially well-populated by
stars whose angular momenta are close to the critical value for capture
$J_{lc}$,
encounters will set up a steady state near the loss cone
in the scattering time corresponding to 
$\theta=\theta_{lc}\approx (r_t/r)^{1/2}$, or
\begin{equation}
T_{lc} \approx {r_t\over r}T_r.
\end{equation}
Figure~\ref{fig:tscales} shows that $T_{lc}$ is shorter than $10^{10}$ 
yr in most galaxies.

But suppose that the loss cone was initially emptied of all stars with
pericenters below some radius $r_0\gg r_t$; 
for instance, a binary black hole might have ejected such
stars via the gravitational slingshot (\S 4).
Once the two black holes had coalesced,
the time to diffusively fill this angular momentum gap would be
\begin{equation}
T_{gap} \approx {r_0\over r}T_r \gg T_{lc}.
\label{eq:tgap}
\end{equation}
In the binary black hole model,
most stars with pericenters
\begin{equation}
r_p \lesssim Ka_h=K{G\mu\over 4\sigma^2}
\label{eq:rp}
\end{equation}
will have been ejected prior to the coalescence.
Here $a_h\gg r_t$ is the semimajor axis of the binary when 
it first becomes ``hard,''
$\mu$ is the reduced mass of the binary,
and $K\approx 2$.
The middle histogram in Figure~\ref{fig:tscales} shows 
$T_{gap}$ for two values of the binary
mass ratio, $m_2/m_1=(0.1,1)$.
In every galaxy,
\begin{equation}
T_{lc}\ll T_{gap} < T_r.
\end{equation}
Whereas $T_{lc}$ is generally shorter than $10^{10}$ yr,
$T_{gap}$ exceeds $10^{10}$ yr in most galaxies
and this is true for all ``core'' galaxies which show
evidence of cusp destruction.

A more careful calculation \cite{wang-05}, 
applying the time-dependent
Fokker-Planck equation to the galaxies in Figure~\ref{fig:ndot},
shows that the time $t_{1/2}$ for the feeding rate
to reach $1/2$ of its steady-state value after loss cone
depletion by a binary black hole is
approximately
\begin{equation}
{t_{1/2}\over 10^{11}{\rm yr}} 
\approx {\mu\over 10^7M_\bullet} \approx 10{q\over\left(1+q\right)^2}
{M_\bullet\over 10^8\msun}
\end{equation}
with $q=m_2/m_1$ the binary mass ratio.
While this is a highly idealized model, it gives an indication
of the time required for loss cone feeding to reach a steady
state in galactic nuclei, and suggests that the tidal
flaring rates in luminous ``core'' galaxies might be much lower
than suggested by Figure~\ref{fig:ndot}.

\section{Interaction of Binary Black Holes with Stars}

Binary supermassive black holes are inevitable consequences
of galaxy mergers.
Unless one of the black holes is very small,
the two quickly form a bound pair at a separation
$a\approx a_h\equiv G\mu/4\sigma^2\approx 
[\mu/4(m_1+m_2)]r_h$ with $\mu$ the reduced mass.
Subsequently the binary separation can decrease, 
but only if the binary is able to exchange
angular momentum with stars or gas.
(See \cite{mm-05} for a review of binary-gas interactions.)
Stars that 
pass within a distance $\sim 3a$ of the binary
undergo a complex 3-body interaction 
followed by ejection of the star at a velocity of order $\mu/m_{12}$
times the binary orbital velocity (the ``gravitational
slingshot'' \cite{sva-74}).
Each ejected star carries away energy and angular momentum,
causing the semi-major axis, eccentricity, orientation, and center-of-mass
velocity of the binary to change and the local density of stars
to drop \cite{hf-80,hills-83,hills-92,mv-92,quinlan-96,merritt-01,merritt-02}.

\begin{figure}[h]
\begin{center}
\includegraphics[width=0.85\textwidth]{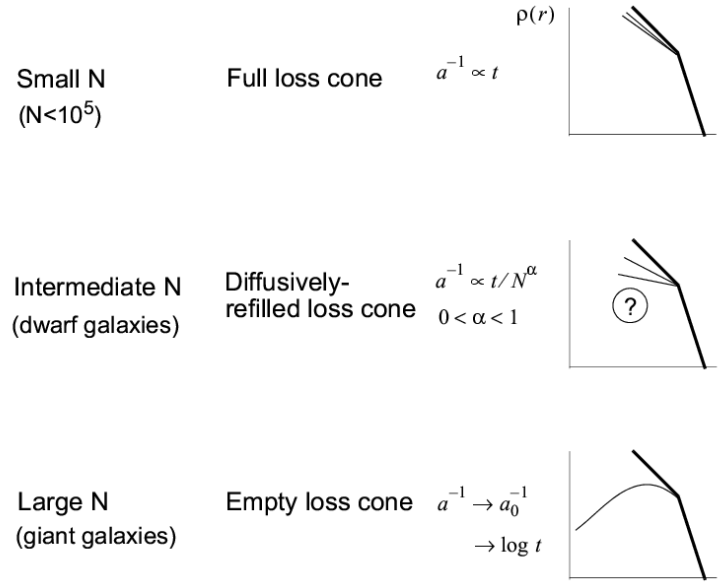}
\end{center}
\caption[]{Physical regimes for the decay of massive black hole
binaries \cite{mm-03}.  \label{fig:regimes}
}
\end{figure}

In principle, $N$-body studies can reveal both the long-term 
evolution of the binary as it interacts with stars, and
the back-reaction of the binary on its stellar surroundings.
But unless great care is taken, $N$-body simulations
are likely to give misleading results.
The reason is that time scales for 
two-body scattering of stars into the binary's loss cone
are shorter by factors of $\sim N/10^{11}$ in the simulations
than in real galaxies, and the
long-term evolution of the binary is likely to be 
dominated by spurious loss cone refilling,
wandering of the binary,
and other noise-driven effects.
Values of $N$ that are easily accessible to direct-summation
$N$-body simulation,
$N\lesssim 10^5$, are so small that stars tend to be scattered
into the binary's loss cone at a faster rate than they are 
kicked out by the gravitational slingshot \cite{mm-03}.
This is analogous to the ``full loss cone'' around a single
black hole and it guarantees that the
binary will never run out of stars.

$N$-body studies
are most reliable when characterizing the
early stages of binary formation and decay. 
Due to algorithmic limitations,
 most such studies 
\cite{ebisuzaki-91,makino-93,makino-97,hss-02,chatterjee-03,makino-04}
have been based on galaxy models with unphysically
large cores.
The first simulations with realistically dense initial conditions
\cite{cruz-01,mm-01,mmvj-02} 
showed that the stellar density around the binary 
drops very quickly after the binary first becomes hard,
converting a $\rho\propto r^{-2}$ cusp to a shallower
$\sim r^{-1}$ cusp.

\begin{figure}[h]
\begin{center}
\includegraphics[width=.75\textwidth]{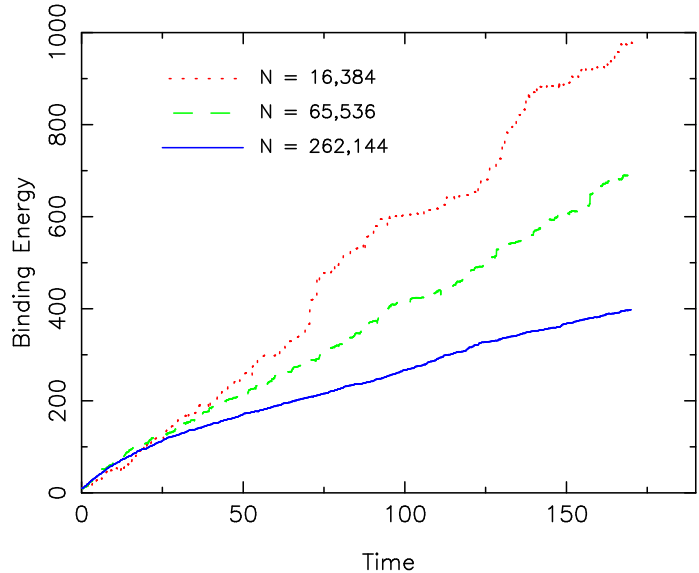}
\end{center}
\caption[]{$N$-dependence of the decay rate of a massive
  binary in $N$-body simulations \cite{szell-04}.
  Each of the three runs started from the same initial state,
  an equal-mass binary with a binding energy of $10$. 
    For small $N$, the long-term decay satisfies $a^{-1}\sim t$,
  the expected behavior in the full-loss-cone regime
  (Fig.~\ref{fig:regimes}).
  As $N$ increases, the lower degree of collisionality
  allows the loss cone around the binary to remain partially
  unfilled and decay slows.
  \label{fig:szell}
}
\end{figure}

But the long-term evolution of massive binaries in real galaxies,
and the effect of binary evolution on nuclear structure, are 
not well understood.
Figure~\ref{fig:regimes} summarizes the likely regimes.
For small $N$, of the order easily accessible to $N$-body
simulations, collisionality is so high that the binary's loss
cone remains full. The binary's binding energy increases linearly
with time, $a^{-1}\propto t$, and the stellar density profile
becomes gradually flatter.
The same sort of evolution would occur in a triaxial nucleus where
centrophilic orbits can continuously repopulate the loss 
cone \cite{poon-04}.
At the other extreme, in giant galaxies, 
relaxation times are very long
and almost no stars would be scattered into the binary's loss cone.
The decay stalls, leaving a ``hole'' in the stellar density \cite{zier-01}.
A mechanism that might contribute to continued decay in this regime
is re-ejection: stars ejected with less than galactic escape velocity
can continue to interact with the binary \cite{mm-03}.
The most interesting, and least-understood, regime is the intermediate
one, 
corresponding to galaxies in which
two-body scattering contributes substantially to the
binary's decay without completely re-filling its loss cone \cite{yu-02}.
Reproducing this regime requires either $N$-body simulations
with $N\gtrsim 10^7$, or a hybrid scheme that combines
a Fokker-Planck integrator with a central binary.
Figure~\ref{fig:szell} shows a step in this direction:
$N$-body integrations
for $N$ up to $0.25\times 10^6$
using a GRAPE-6 special-purpose computer, coupled with
a chain regularization algorithm
to follow close interactions of star and black hole particles
\cite{szell-04}.
The $N$-dependence of the decay rate is clearly visible.
Similar results were obtained \cite{makino-04} using a 
lower-accuracy $N$-body code.
Increasing $N$ by an order of magnitude will soon be feasible
and this will allow simulations in which the binary's 
evolution is dominated by slow diffusion into the loss cone,
as in real galaxies.

\section{Brownian Motion}

Both single and binary black holes undergo a random walk in
momentum space as their motion is perturbed by gravitational
encounters with nearby stars.
Simple encounter theory predicts that a massive black hole
reaches a state of kinetic energy
equipartition with the stars, giving it a
mean square velocity of
$\langle V^2\rangle\approx 3(m/M_\bullet)\sigma^2$.
This ``gravitational Brownian motion'' has been suggested
as a mechanism for enhancing the supply of stars to
a massive binary; 
indeed it has been asserted \cite{quinlan-97,chatterjee-03} that
wandering of the binary's center of mass can itself 
guarantee a continued supply of stars.

While Brownian motion probably does affect the decay rate of
binaries in the $N$-body simulations \cite{mm-03}, it is doubtful
that the effect is significant in real galaxies.
The Brownian velocity of {\it single} black holes is found 
in $N$-body simulations to be
\cite{laun-04}
\begin{equation}
{1\over 2}M_\bullet\langle V^2\rangle \approx {3\over 2}m\tilde\sigma^2
\label{eq:brown}
\end{equation}
where $\tilde\sigma^2$ is the 1D, mean square stellar velocity within
a region $r\lesssim 0.5 r_h$ around the black hole and $m$
is the stellar mass.
Equation (\ref{eq:brown}) holds in galaxy models with
a wide range of nuclear density slopes, black hole masses
and particle numbers up to $10^6$, the largest values so 
far used in direct-summation $N$-body simulations \cite{dorband-03}.
In the case of the Milky Way black hole, equation (\ref{eq:brown})
implies $V_{\rm rms}\approx 0.17$ km s$^{-1}$ (assuming $m=M_\odot$),
a little larger than the ``equipartition'' value.
The reason is that the black hole responds to perturbations from
stars whose
velocities have themselves been increased by the presence
of the black hole \cite{laun-04}.

Brownian motion of a massive {\it binary} is larger than
that of a single black hole, for two reasons \cite{merritt-01}. 
(1) Stars are ejected superelastically from the binary, imparting 
a greater momentum to the binary than they would to a single
black hole. (2) The dynamical friction force acting on the
binary is less than that acting on a single particle
due to the randomization of the ejection angles.
The first factor dominates and produces an extra fractional contribution
to $\langle V^2\rangle$ that is approximately $H/(32\sqrt{2}\pi\ln\Lambda)$,
with $H\approx 16$ the dimensionless hardening rate of the binary
\cite{quinlan-96}.
Thus the Brownian motion
of a binary is only slightly larger than that of a single
black hole of the same mass.
This prediction has been verified in $N$-body simulations 
\cite{mm-01,makino-04}.
The rms displacement of a binary from its otherwise central
location would be very small in a real galaxy, probably even
less than the separation between the two black holes.

The claim 
that wandering can guarantee a continued supply
of stars to a binary
is based on the following conceptual model
\cite{quinlan-97,chatterjee-03}.
As $N$ increases,
the wandering first drops as expected from equation (\ref{eq:brown}),
but at sufficiently large $M_\bullet/m$ the binary 
empties its loss cone, producing a low-density core 
in which the binary is easily displaced.
The amplitude of the binary's center-of-mass 
motion increases, allowing it to
interact with a larger number of stars.

The motivation for this model was the complex
$N$-dependence observed in a set of binary
decay simulations \cite{quinlan-97,chatterjee-03}:
the decay rate first dropped with increasing $N$ until
$N\approx 2\times 10^5$, then remained constant when
$N$ was doubled.
But the postulated $N$-dependence of the {\it wandering} was
never verified.
Even if the interpretation in terms of wandering is correct,
extrapolating the $N$-body results from 
$N\lesssim 10^6$ to $N\approx 10^{11}$ is problematic.
Furthermore the galaxy models were Plummer spheres which have
a much lower degree of concentration than real galaxies.
and the wandering amplitudes inferred by these authors
would be enormous if scaled to real galaxies, $\gtrsim 10^2$ pc,
Extended simulations of binary hardening in realistic galaxy models
with $N>10^6$ should soon settle this matter.

\section{Evidence for Cusp Destruction}

\begin{figure}[h]
\begin{center}
\includegraphics[width=.80\textwidth]{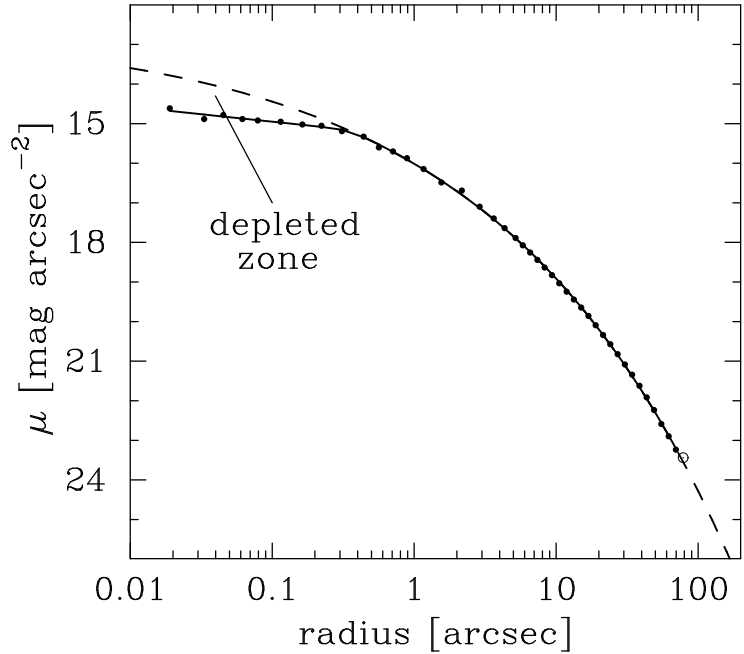}
\end{center}
\caption[]{Observed surface brightness profile of NGC 3348.
  The dashed line is a Sersic model fit to the large-radius
  data.  The mass deficit is illustrated by the area
  designated ``depleted zone'' and the corresponding mass is
  roughly $3\times 10^8\msun$ \cite{graham-04}.}
  \label{fig:sersic}
\end{figure}

The structural parameters of elliptical galaxies display
a smooth dependence on galaxy luminosity from
dE galaxies up to the brighter ellipticals,
$M_V\approx -20$ \cite{graham-03}.
Above this luminosity, galaxies generally have cores, 
regions near the
center where the density falls below the inward
extrapolation of the outer profiles 
(although the density almost always continues to rise inward).
It is natural to attribute the cores to stellar ejection
by a binary black hole and to use the
``mass deficit'' as a quantitative test of the binary 
black hole model (Fig.~\ref{fig:sersic}).
In two studies \cite{milos-02,ravin-02},
pre-merger profiles for a sample of galaxies
were constructed by extrapolating
power laws, $\rho\propto r^{-\gamma_0}$, inward of the
core radius.
Milosavljevic et al. \cite{milos-02} found mean mass
deficits of $\sim 1M_\bullet$ for $\gamma_0=1.5$.
More recently, Graham \cite{graham-04} 
fit the outer luminosity data to Sersic
profiles and defined the mass deficit with respect
to this template; mass deficits were found to be slightly larger,
about twice $M_\bullet$.\footnote[1]{Graham (2004) presents 
his mass deficits as significantly
{\it smaller} than in the earlier study, but he
bases his comparison exclusively on values of $M_{\rm def}$ computed
by Milosavljevic et al. using $\gamma_0=2$; 
the latter authors computed $M_{\rm def}$ for a range of values 
$1.5\le\gamma_0\le 2.0$.}
These numbers are reasonable but it is difficult to
say more given the difficulties cited above in interpreting
$N$-body simulations of binary evolution.

An interesting unsolved problem is the effect of multiple
mergers on the central density profiles of galaxies 
containing black holes.
A simple argument suggests that the net effect,
e.g. the core size, should increase both with the final
black hole mass {\it and} with the number of merger events.
Consider the merger of two galaxies with steep central
cusps and black holes of mass $m_1$ and $m_2$; the result
will be a galaxy with a shallower central profile
and black hole of mass $m_{12}=m_1+m_2$.
Now imagine that that same galaxy is produced by the
merger of two galaxies whose central density cusps had
already been destroyed by previous merger events. 
The final merger would lower the central density still more,
producing a larger core for the same $m_{12}$.
Testing this prediction via $N$-body simulations will
be difficult for all the reasons discussed above,
but it should be possible at least to test how
the early evolution of binaries affects the central
profiles of galaxies in multiple mergers.

The idea that cores grow cumulatively in mergers
was used by Volonteri et al. \cite{volonteri-03} to predict the
evolution of luminous and dark matter cores
in galaxies.
In their most extreme (``core preservation'') model,
the binaries were assumed to eject as much stellar mass 
as needed to shrink to a separation where gravitational radiation
losses induce coalescence.
The result, after $10^{10}$ yr, was cores with sizes of order $3r_h$.
While intriguing, this model lacks a firm foundation in
the $N$-body simulations.

In small dense galaxies, a destroyed cusp would be expected to
re-form via the collisional mechanism discussed above,
{\it if} the relaxation time at the core radius is
less than the elapsed time since the merger.
Alternatively, steep cusps may be due to star formation
that occurred after the most recent merger.

There is one context in which formation of cores
by black holes has been tested and shown to work.
Globular clusters in the Large and Small
Magellanic Clouds exhibit a range of ages and core sizes,
and the mean core radius increases as the logarithm
of the age \cite{elson-91}.
The observed age dependence is reproduced very well by a model
in which the first population of stars leave behind $\sim 10M_\odot$
remnants which sink to the center and displace the less massive stars before
ejecting themselves via three-body interactions \cite{merritt-04b}.
This model may have some relevance to galactic nuclei
if supermassive binaries stall long enough for third or fourth black
holes to fall in.

\begin{figure}[h]
\begin{center}
\includegraphics[width=.80\textwidth]{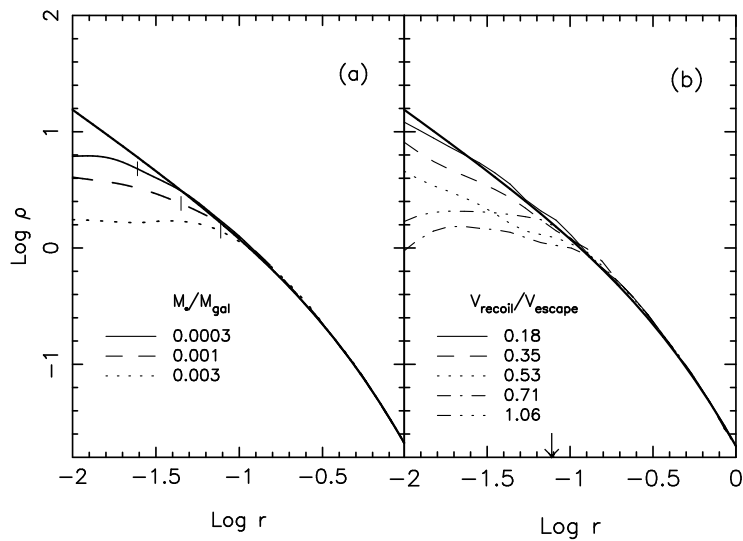}
\end{center}
\caption[]{Effect on the nuclear density profile of 
black hole ejection.
The initial galaxy model (black line) has a $\rho\sim r^{-1}$
density cusp.
(a) Impulsive removal of the black hole.
Tick marks show the radius of the black hole's sphere of influence
$r_h$ before ejection.
A core forms with radius $\sim 2r_h$.
(b) Ejection at velocities less than escape velocity.
The black hole has mass $0.003 M_{\rm gal}$;
the galaxy is initially spherical and the black hole's
orbit remains nearly radial as it decays via dynamical friction.
The arrow marks $r_h$.
}
  \label{fig:change}
\end{figure}

\section{Black Hole Ejections}

If two black holes do manage to coalesce due to emission
of gravitational radiation, linear momentum carried
by the waves will impart a kick to the coalesced hole
of amplitude $\sim 10^{-3}c$ \cite{fitchett-83}.
The recoil velocity depends in a complicated way on
the mass ratio and spins of the two black holes and
on the spin orientations \cite{favata-04}.
At present only the mass ratio dependence is well 
understood: the recoil velocity peaks at $m_2/m_1\approx 0.38$
and falls to zero at $m_2/m_1=0$ or $1$.
As discussed by Hughes, Favata \& Holz in these proceedings,
uncertainty about the strong-field behavior of
the effect allows only plausible upper and lower limits
to be placed on the magnitude of the kicks.
However for moderately large spins and prograde capture,
even the lower limits exceed $100$ km s$^{-1}$ for 
$0.2\lesssim m_2/m_1 \lesssim 0.6$,
and the upper limit estimates reach $\sim 500$ km s$^{-1}$
for favorable mass ratios \cite{mmfhh-04}.

An important possible consequence of the kicks is ejection
of black holes from galaxies, particularly in the early
universe when potential wells were shallow 
\cite{mmfhh-04,haiman-04,yoo-04}.
Whether or not a black hole is completely ejected,
displacement of the black hole transfers energy to the nucleus
and lowers its density within a region of size $\sim r_h$
(Fig.~\ref{fig:change}). 
Impulsive removal at $V_{\rm kick}\gg V_{\rm escape}$
produces a core of roughly constant density within
a radius $\sim 2r_h$ if the initial density profile is 
$\rho\propto r^{-1}$.
The net effect is greater if 
$V_{\rm kick}\lesssim V_{\rm escape}$ since the black hole
can return to the nucleus several times before settling down.
For $V_{\rm kick}\lesssim 0.25 V_{\rm escape}$ the change
in the density is negligible.
Cores formed in this way have a similar size to those produced
in the binary black hole model and complicate the interpretation
of mass deficits.

\end{document}